# Induction and Amplification of Non-Newtonian Gravitational Fields

M. Tajmar[*]

*Austrian Research Centers Seibersdorf, A-2444 Seibersdorf, Austria*

C. J. de Matos[†]

*ESA-ESTEC, Directorate of Scientific Programmes, PO Box 299, NL-2200 AG Noordwijk, The Netherlands*

**Abstract**

One obtains a Maxwell-like structure of gravitation by applying the weak-field approximation to the well accepted theory of general relativity or by extending Newton's laws to time-dependent systems. This splits gravity in two parts, namely a gravitoelectric and gravitomagnetic (or cogravitational) one. Both solutions differ usually only in the definition of the speed of propagation, the lorentz force law and the expression of the gravitomagnetic potential energy. However, only by extending Newton's laws we obtain a set of Maxwell-like equations which are perfectly isomorphic to electromagnetism. Applying this theory to explain the measured advance of the mercury perihelion we obtain exactly the same prediction as starting from general relativity theory. This is not possible using the weak-field approximation approach. Due to the obtained similar structure between gravitation and electromagnetism, one can express one field by the other one using a coupling constant depending on the mass to charge ratio of the field source. This leads to equations e.g. of how to obtain non-Newtonian gravitational fields by time-varying magnetic fields. Unfortunately the coupling constant is so small that using present day technology engineering applications for gravitation using electromagnetic fields are very difficult. Calculations of induced gravitational fields using state-of-the-art fusion plasmas reach only accelerator threshold values for laboratory testing. Possible amplification mechanisms are mentioned in the literature and need to be explored. We review work by Henry Wallace suggesting a very high gravitomagnetic susceptibility of nuclear half-spin material as well as coupling of charge and mass as shown by e.g. torque pendulum experiments. The possibility of using the principle of equivalence in the weak field approximation to induce non-Newtonian gravitational fields and the influence of electric charge on the free fall of bodies are also investigated, leading to some additional experimental recommendations.

---

[*] Research Scientist, Space Propulsion, Phone: +43-50550-3142, Fax: +43-50550-3366, E-mail: martin.tajmar@arcs.ac.at
[†] Staff Member, Science Management Communication Division, Phone: +31-71-565 3460, Fax: +31-71-565 4101, E-mail: clovis.de.matos@esa.int

**Introduction**

The control and modification of gravitational fields is a dream pursued by propulsion engineers and physicists around the world. NASA's Breakthrough Propulsion Physics Project is funding exploratory research in this area to stimulate possible breakthroughs in physics that could drastically lower costs for access to space[1]. Although not commonly known, Einstein's well accepted general relativity theory, which describes gravitation in our macroscopic world, allows induction phenomena of non-Newtonian gravitational fields similar to Faraday induction in electromagnetic fields by moving heavy masses at high velocities.

The basis for such phenomena are even dating back before general relativity theory when Oliver Heaviside[2] in 1893 investigated how energy is propagated in a gravitational field. Since energy propagation in electromagnetic fields is defined by the Poynting vector – a vector product between electric and magnetic fields – Heaviside proposed a gravitational analogue to the magnetic field. Moreover he postulated that this energy must also be propagating at the speed of light. Another approach to the magnetic part of gravity is to start from Newtonian gravity and add the necessary components to conserve momentum and energy[3]. This leads to the same magnetic component and a finite speed of propagation, the speed of light.

Heaviside's gravitomagnetic fields are hidden in Einstein's Tensor equations. Alternatively, general relativity theory can be written as linear perturbations of Minkowski spacetime. Forward[4] was the first to show that these perturbations can be rearranged to assemble a Maxwell-type structure which splits gravitation into a gravitoelectric (classical Newtonian gravitation) and a gravitomagnetic (Heaviside's prediction) field. The magnetic effects in gravitation are more commonly known as the Lense-Thirring or frame dragging effect describing precision forces of rotating masses orbiting each other. NASA's mission Gravity Probe B will look for experimental evidence of this effect. Similar to electrodynamics, a variation in gravitomagnetic fields induces a gravitoelectric (non-Newtonian) field and hence provides the possibility to modify gravitation.

Since both gravitation and electromagnetism have the same source, the particle, the authors recently published a paper evaluating coupling constants between both fields[5] based on the charge-to-mass ratio of the source particle. This paper will review the coupling between gravitation and electromagnetism and point out the limits of present day technology and the expected order of magnitude of non-Newtonian gravitational fields that can be created by this method. Possible amplification mechanism such as ferro-gravitomagnetism and more speculative work published in the literature will be reviewed.

The principle of equivalence in the limit of weak gravitational fields (the gravitational Larmor theorem) will be explored and a possible new effect (the gravitomagnetic Barnett effect) recently suggested by the authors is discussed[6]. However the direct detection of this effect is pending on the possibility to have materials with high gravitomagnetic susceptibility. Nevertheless we show that the principle of equivalence in the weak field approximation together with the gravitational Poynting vector associated with induced non Newtonian gravitational fields (through angular acceleration) account properly for the conservation of energy in the case of cylindrical mass with angular acceleration. This is an encouraging result regarding the possible detection of macroscopic non-Newtonian gravitational fields induced through the angular acceleration of the cylinder in the region located outside the rotating cylinder. The detection of these non-Newtonian gravitational fields outside the cylinder would represent an indirect evidence of the existence of the gravitomagnetic Barnett effect.

Finally the free fall of a massive cylinder carrying electric charge is studied. It is shown that in order to comply with the law of conservation of energy, and with the equivalence principle, the acceleration with which the cylinder will fall depends on its electric charge, its mass and its length.

If the last two effects exposed above are experimentally detected, a technology that can control the free fall of bodies with mass in the laboratory is at hand. If the result is negative, a better empirical understanding of Einstein's general relativity theory in the limit of weak gravitational fields and when extended to electrically charged bodies, would have been achieved, which is a significant scientific result as well.

## Maxwell Structure of General Relativity Theory

Einstein's field equation[7] is given by

$$R_{ab} - \frac{1}{2} g_{ab} R = \frac{8 p G}{c^4} T_{ab} \qquad (1)$$

During the linearization process, the following limitations are applied:

1. all motions are much slower than the speed of light to neglect special relativity
2. the kinetic or potential energy of all bodies being considered is much smaller than their mass energy to neglect space curvature effects
3. the gravitational fields are always weak enough so that superposition is valid
4. the distance between objects is not so large that we have to take retardation into account

We therefore approximate the metric by

$$g_{ab} \cong \boldsymbol{h}_{ab} + h_{ab} \qquad (2)$$

where the greek indices $\boldsymbol{a}, \boldsymbol{b} = 0, 1, 2, 3$ and $\boldsymbol{h}_{ab} = (+1, -1, -1, -1)$ is the flat spacetime metric tensor, and $|h_{ab}| \ll 1$ is the perturbation to the flat metric. By proper substitutions and after some lengthy calculations[5] (the reader is referred to the literature for details), we obtain a Maxwell structure of gravitation which is very similar to electromagnetics and only differs due to the fact that masses attract each other and similar charges repel:

$$div\ \vec{E} = \frac{\rho}{\epsilon_0} \qquad\qquad div\ \vec{g} = -\frac{\rho_m}{\epsilon_g}$$

$$div\ \vec{B} = 0 \qquad\qquad div\ \vec{B}_g = 0$$

$$rot\ \vec{E} = -\frac{\partial \vec{B}}{\partial t} \qquad\qquad rot\ \vec{g} = -\frac{\partial \vec{B}_g}{\partial t} \qquad (3)$$

$$rot\ \vec{B} = \mu_0 \rho \vec{v} + \frac{1}{c^2}\frac{\partial \vec{E}}{\partial t} \qquad rot\ \vec{B}_g = -\mu_g \rho_m \vec{v} + \frac{1}{c^2}\frac{\partial \vec{g}}{\partial t}$$

<div align="center">Maxwell Equations      Maxwell-Einstein Equations

(Electromagnetism)      (Gravitation)</div>

where $\vec{g}$ is the gravitoelectric (or Newtonian gravitational) field and $\vec{B}_g$ the gravitomagnetic field. The gravitational permittivity $\epsilon_g$ and gravitomagnetic permeability $\mu_g$ is defined as:

$$\epsilon_g = \frac{1}{4\pi G} = 1.19 \times 10^9\ \frac{\text{kg} \cdot \text{s}^2}{\text{m}^3} \qquad (4)$$

$$\mu_g = \frac{4\pi G}{c^2} = 9.31 \times 10^{-27}\ \frac{\text{m}}{\text{kg}} \qquad (5)$$

by assuming that gravitation propagates at the speed of light $c$. Although not unusual, this assumption turns out to be very important. Only if gravity propagates at $c$ the Maxwell-Einstein equations match the ones obtained from adding necessary terms to Newtonian gravity to conserve momentum and energy[3]. Moreover, the authors could show that with this set of equations, the advance of the Mercury perihelion – one of the most successful tests of general relativity – can be calculated giving the exact prediction than without linearization[8]. This is a very surprising result because the advance of Mercury's perihelion is attributed to a space curvature in general relativity (Schwarzschild metric) which we neglected in our linearization process. The assumption of $c$ as the speed of gravity propagation also implies that the Lorentz force law and the gravitomagnetic potential energy differ from their electromagnetic counterparts by a factor of four[8]. Therefore some authors[4] use $c/2$ as the speed of gravity propagation to get a gravity Lorentz force law similar to electromagnetics.

The Einstein-Maxwell equations allow to clearly see the gravitomagnetic component of gravitation and the possibility to induce non-Newtonian gravitational fields. Their close relation to electrodynamics allow to transform electromagnetic calculations into their gravitational counterparts[9].

## Coupling of Electromagnetism and Gravitation in General Relativity

By comparing gravitation and electromagnetism in Equation (3), we see that both fields are coupled by the e/m ratio of the field source and we can write:

$$\begin{aligned} \vec{g} &= \mathbf{k} \cdot \vec{E} \\ \vec{B}_g &= \mathbf{k} \cdot \vec{B} \end{aligned} \quad (6)$$

using the coupling coefficient **k**

$$\mathbf{k} = -\frac{m}{e}\frac{\mathbf{m}_g}{\mathbf{m}_0} = -\frac{m}{e}\frac{\mathbf{e}_0}{\mathbf{e}_g} = -7.41 \times 10^{-21} \cdot \frac{m}{e} \quad (7)$$

Obviously, this coefficient is very small and gravitational effects associated with electromagnetism have never been detected so far[10]. By combining Equation (6) with Equation (3), we see how both fields can induce each other:

$$\begin{aligned} rot\ \vec{E} &= -\frac{1}{\mathbf{k}}\frac{\partial \vec{B}_g}{\partial t} \\ rot\ \vec{B} &= \frac{e}{m}\mathbf{m}_0\, r_m \vec{v} + \frac{1}{\mathbf{k}}\frac{1}{c^2}\frac{\partial \vec{g}}{\partial t} \end{aligned} \quad (8)$$

Coupled Maxwell-Einstein Equations

(Gravitation→Electromagnetism)

$$\begin{aligned} rot\ \vec{g} &= -\mathbf{k}\frac{\partial \vec{B}}{\partial t} \\ rot\ \vec{B}_g &= -\frac{m}{e}\mathbf{m}_g\, r\vec{v} - \mathbf{k}\frac{1}{c^2}\frac{\partial \vec{E}}{\partial t} \end{aligned} \quad (9)$$

Coupled Maxwell-Einstein Equations

(Electromagnetism→Gravitation)

For an electron in a vacuum environment **k**=4.22x10$^{-32}$ kg/C. For example, let us consider an infinitely long coil as shown in Figure 1.

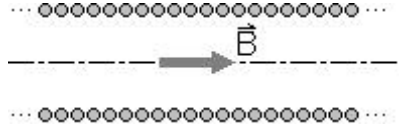

**Figure 1   Magnetic Field Induced in a Coil**

The magnetic field induced in the center line is then

$$B = \mu_0 I n \qquad (10)$$

where *I* is the current and *n* is the number of coil wounds per length unit. For a current of 10,000 Ampére and 1,000 wounds per meter, the magnetic field would be $B$=12.56 T which is state of the art. The corresponding gravitomagnetic field is then $B_g$=5.3x10$^{-31}$ s$^{-1}$. Even using a coil with 100,000 wounds to induce an electric field, the amplitude of the resulting gravitational field would only be in the order of $g$=10$^{-26}$ ms$^{-2}$. This is much too small to be detected by any accelerometers having measurement thresholds of 10$^{-9}$ ms$^{-2}$. By using heavy ions in a plasma instead of electrons we can increase the m/e ratio by 6 orders of magnitude, however, the magnetic fields to contain such a plasma transmitting a similar current of 10,000 Ampére are out of reach.

Nevertheless, although the induced gravitational fields are very small, in principle it is possible to create non-Newtonian gravitational fields along the same principles as we are used to in electromagnetism.

**Amplification Mechanisms**

Since all these electromagnetic-gravitational phenomena are so small, how can we amplify the coupling coefficient in order to obtain measurable non-Newtonian fields?

*Gravitation-Magnetism*

Similar to para-, dia-, and ferro-magnetism, the angular and spin momentums from free electrons in material media could be used to obtain a gravitomagnetic relative permeability $\mu_{gr}$ which increases the gravitomagnetic field $\vec{B}_g$. Since an alignment of magnetic moments causes also an alignement of gravitomagnetic moments, the gravitomagnetic susceptibility will be the same as the magnetic susceptibility in a magnetized material[5]

$$c_g = c \qquad (11)$$

For our example of the coil in Figure 1, a ferromagnetic core would accordingly increase the gravitomagnetic field and induced non-Newtonian gravitational field by three orders of magnitude. Although significant, the resulting fields are still too low to be detected.

*Coupling of Charge and Mass*

All our discussions up to now are based on a coupling at the source particle by the e/m ratio. However, an additional coupling between charge and mass of the source itself might exist and provide a significant amplification mechanism.

Well accepted peer-review journals like Nature and Foundations of Physics featured articles on this topic describing experiments that suggest a coupling between charge and mass in combination with rotation (or acceleration, movement in general). Dr. Erwin Saxl published an article[11] reporting a period change of a torque pendulum if the pendulum was charged. A positive charge caused the pendulum to rotate slower than when it was charged negatively, Figure 3 shows his observations with a small asymmetry of the period change between positive and negative potentials applied to the pendulum. The period is expressed by

$$T = \text{Constant} \cdot \sqrt{\frac{m}{g}} \qquad (12)$$

where *m* is the mass of the pendulum and *g* the Earth's gravitational acceleration. Assuming that *g* is not changed (it is highly improbable that the whole Earth is affected), Saxl's measurement can be interpreted as a change of the pendulum's mass by applying an electric potential to it.

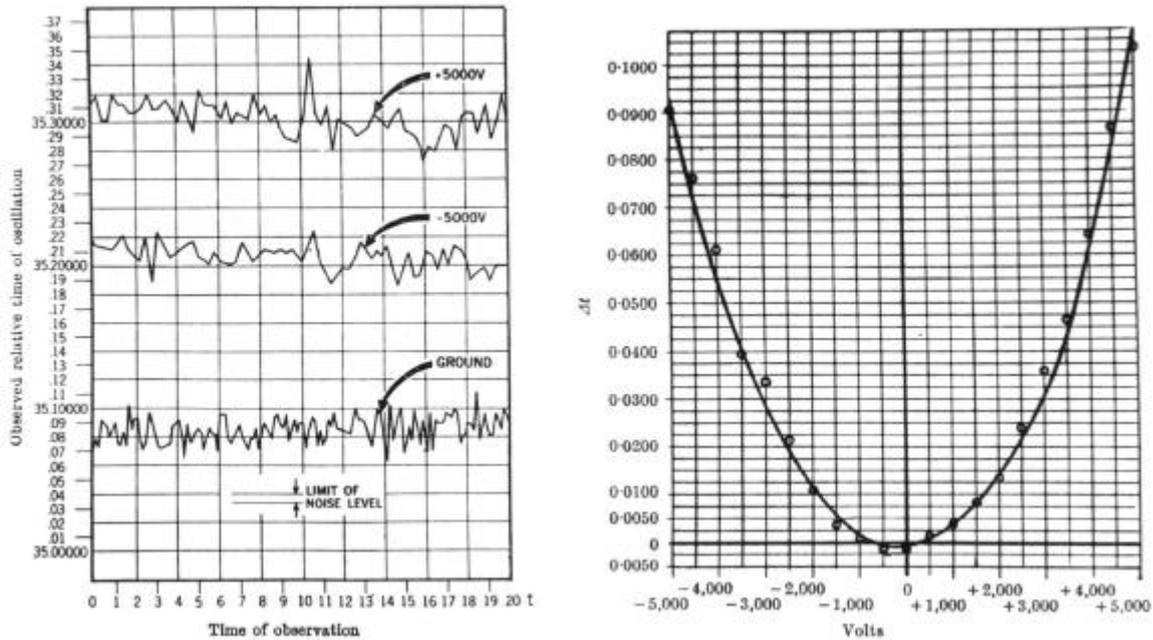

**Figure 3   Change of Torque Pendulum Period vs. Applied Potential[11]**

Prof. James Woodward from the University of California reported experiments of accelerating masses that, on the other hand, charged up according to their mass and speed of rotation. His experiments were done both for rotating masses[12] as well as for linear accelerated test bodies[13]. Published in the Foundations of Physics and General Relativity and Gravitation, he suggested a broader conservation principle including mass, charge and energy. Results of a test body hitting a target and inducing a charge are shown in Figure 4. His results follow

$$q' \cong \text{Constant} \cdot m \cdot a \tag{13}$$

where $q'$ is the induced charge, $m$ the test mass and $a$ the acceleration (from rotation or calculated from the impact velocity).

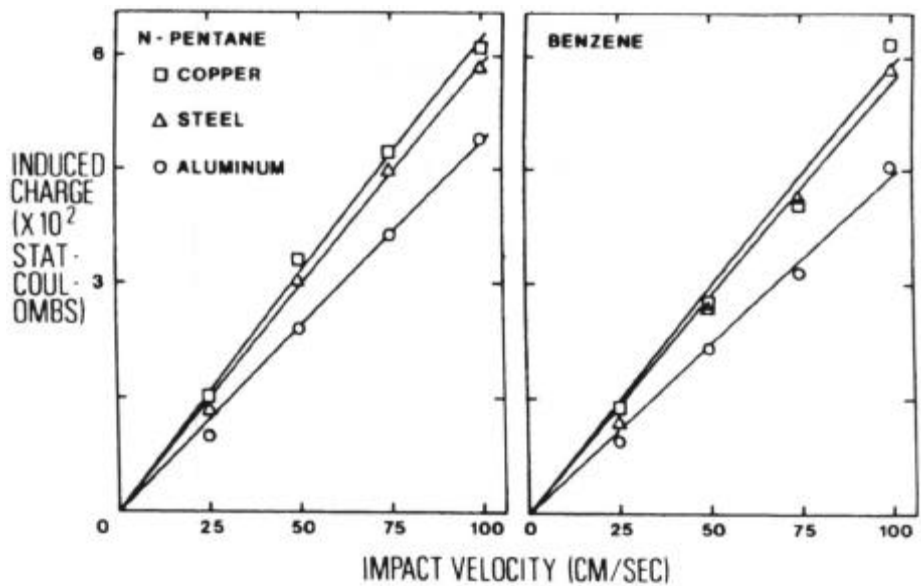

**Figure 4   Charged Induced by Body Hitting Target[13]**

Hence, both Saxl and Woodward experimentally reasoned a relationship between charge, mass and acceleration. A combination of all these factors to reduce/increase the weight of a body is described in a patent by Yamashita and Toyama[14]. A cylinder was rotated and charged using a Van der Graff generator. During operation the weight of the rotating cylinder was monitored on a scale. The setup is shown in Figure 5. If the cylinder was charged positively, a positive change of weight up to 4 grams at top speed was indicated. The same charge negative produced a reduction of weight of about 11 grams (out of 1300 grams total weight). This is an asymmetry similar to the one mentioned by Saxl[11]. Also the relationship between charge, rotation and mass is similar to Saxl and Woodward. The experimentors note that the weight changed according to the speed of the cylinder ruling out electrostatic forces, and that it did not depend on the orientation of rotation ruling out magnetic forces. The reported change of weight (below 1 %) is significant and indicates a very high order of magnitude effect.

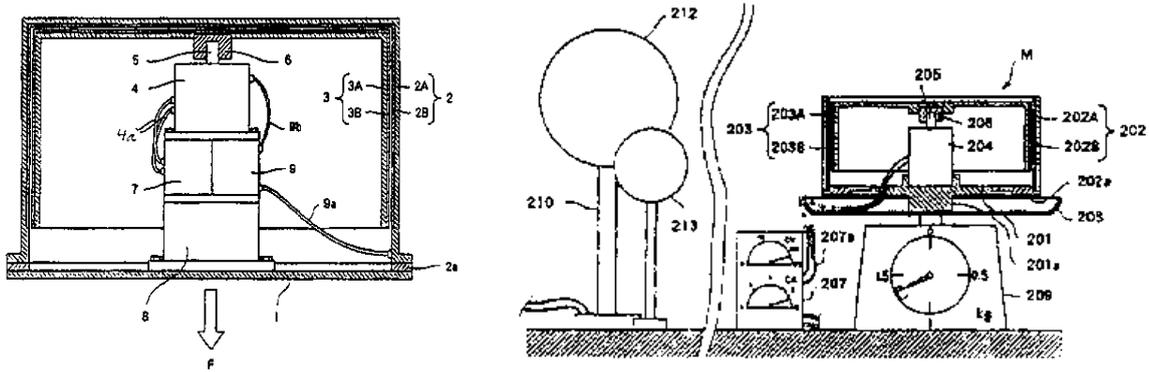

**Figure 5** Setup of Charged Rotating Cylinder on Scale[14]

*Alignment of Nuclear Spins*

Henry Wallace, an engineer from General Electric, holds three patents on a method to produce a macroscopic gravitomagnetic and gravitational field by aligning nuclear spins due to rotation[15-17]. He claims that if materials with a net nuclear half-spin (one neutron more than protons in the nucleus) are rotated, this nuclear spin is aligned and produces a macroscopic gravitational effect. This is in fact similar to the Barnett effect where a metal rod is rotated and magnetisation of the material is observed. However, macroscopic magnetism in electromagnetism is caused by spin alignment of electrons, nuclear magnetism plays a very minor role due to the much higher mass of a proton or neutron compared to the electron. In a gravitational context the difference in mass is no major drawback anymore and nuclear magnetism should be on the same order of magnitude than electron magnetism. Usually, very low temperatures in the order of nano Kelvin are required to align nuclear spins, simple rotation would be much more easy.

The contribution of neutron spins to gravitomagnetic fields is theoretically on the order of ferromagnetism[5]. However, since Wallace claims to have measured at least the induction of nuclear spin alignment in a rotating detector material – by what he thinks a gravitomagnetic field, possible unknown amplification mechanisms (quantum gravity, nuclear strong force interaction) could cause much higher order of magnitude effects.

His setup is shown in Figure 6. A generator assembly (test mass rotating in 2 axis) is mounted on the left side and a detector assembly (similar to generator) is mounted on the right side with the possibility of rotation in the plane of the paper. A laser is monitoring the oscillations of this detector assembly. If both are rotated in the same orientation and counter wise, the laser detected a difference (Figure 7) which Wallace attributed to a force field. Since it only depended on the nuclear spin (e.g. Iron did not work but is a strong ferromagnetic material), Wallace ruled out magnetism as the origin of the force. In a different setup he showed that the field generated could constructively reduce the vibrational degrees of freedom of the crystal structure resulting in a change of its electrical properties (Figure 8).

Hence, there is quite some experimental evidence for an amplification mechanism through nuclear magnetism to generate non-Newtonian gravitational fields using effects predicted by general relativity theory.

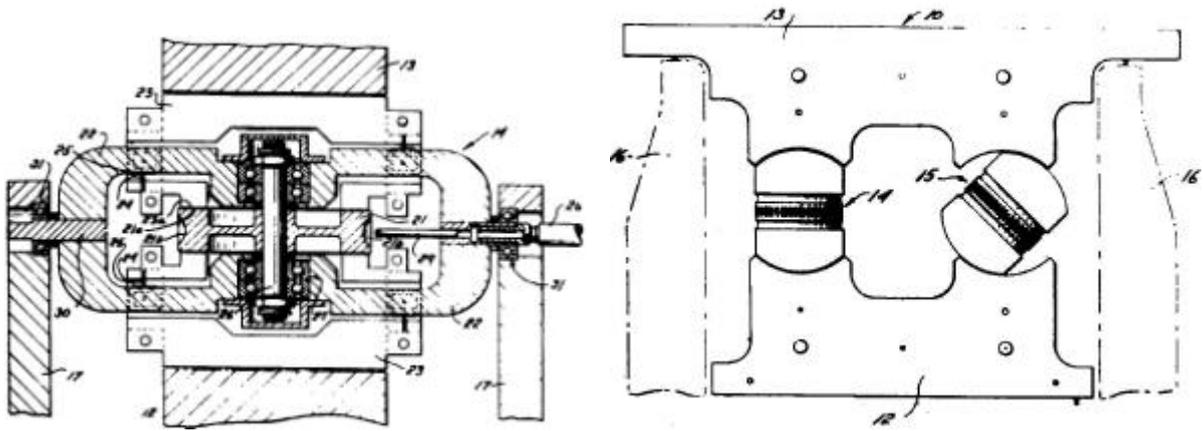

**Figure 6   Setup of Rotating Test Mass (2 Axis) and Generator (Left) and Detector (Right) Position**[15]

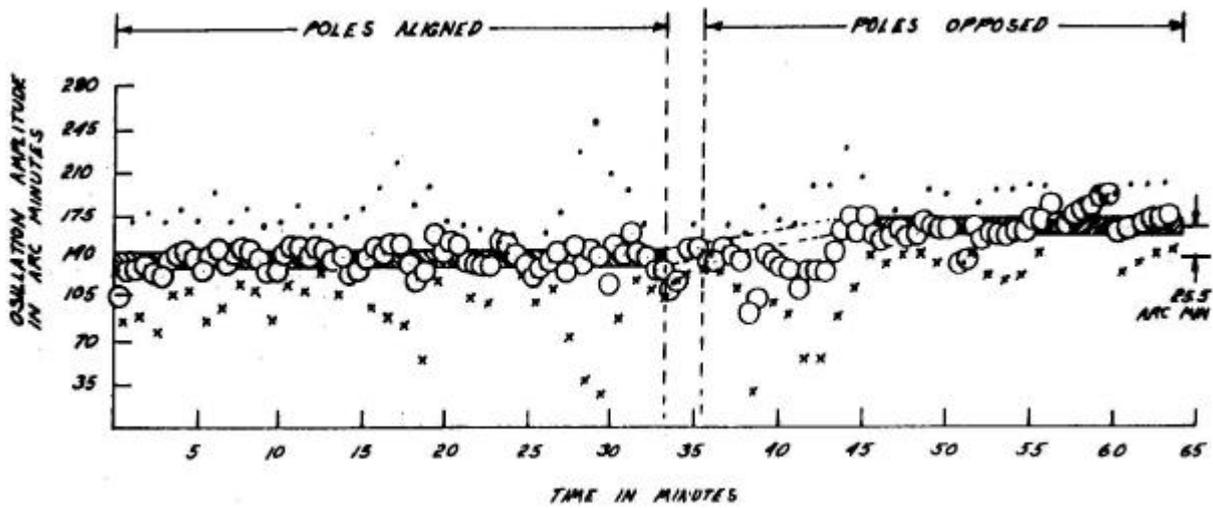

**Figure 7   Oscillations of Detector Assembly**[15]

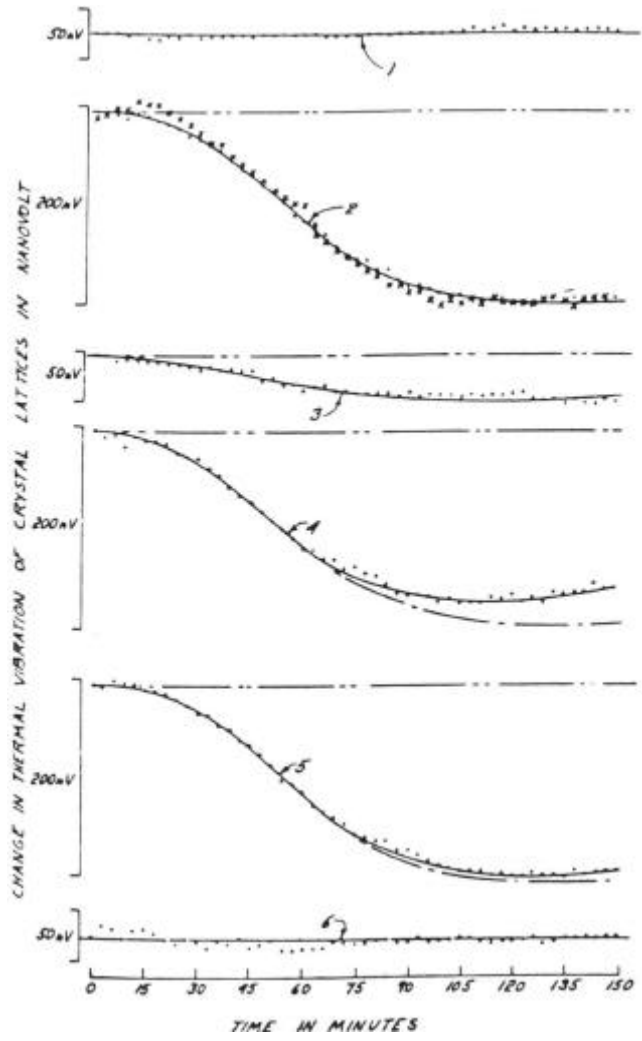

**Figure 8   Change in Thermal Vibration of Crystal Lattices[15]**

## Principle of Equivalence and High-Order of Magnitude Non-Newtonian Gravitation

We explored the limits of inducing non-Newtonian gravitation using general relativity theory as well as looking at possible and speculative amplification mechanisms. Let us go back to the foundation of gravitation itself and explore the principle of equivalence in the limit of weak gravitational fields.

Einstein based his thoughts of gravitation on a famous Gedankenexperiment explaining the principle of equivalence: An observer can not distinguish between being inside a falling elevator or in a uniform gravitational field. Based on this equivalence, he developed the geometrical structure of general relativity. In the limit of weak garvitational fields, this simple Gedankenexperiment however is not complete as it covers only gravitoelectric fields and not the magnetic component of gravitation. According to the Larmor theorem of electromagnetics, a magnetic field can be replaced locally by a rotating reference frame with the Lamor frequency

$$w_L = \frac{1}{2}\frac{e}{m}B \tag{14}$$

The same argument applies for gravitation and a rotating reference frame rotating with the Lamor frequency can replace a gravitomagnetic field

$$w_{Lg} = -B_g \tag{15}$$

independent of the particle mass. The principle of equivalence[18] for weak gravitational fields (neglecting space curvature) also called gravitational Larmor theorem (GLT) should then be:

*An observer can not distinguish between a uniformly accelerated ($\dot{\vec{v}}$) reference frame rotating with the gravitational Larmor frequency ($w_{Lg}$) and a reference frame at rest in a corresponding gravitational field ($\vec{g} = -\dot{\vec{v}}$, $w_{Lg} = -\vec{B}_g$).*

But what happens if the speed of rotation of the elevator changes? According to the GLT, this would correspond to a change of a gravitomagnetic field flux and therefore induce a non-Newtonian gravitational component according to the gravitational Faraday law:

$$e_g = \oint_\Gamma \vec{g}\cdot d\vec{l} = -\frac{d f_{gm}}{dt} = -\frac{d}{dt}\iint_\Sigma \vec{B}_g \cdot d\vec{s} = \frac{d}{dt}\iint_\Sigma \vec{\Omega}\cdot d\vec{s} \tag{16}$$

where $\vec{g}$ is the non-Newtonian gravitational field, $\Gamma$ and $\Sigma$ are respectively the contour and surface of integration, $f_{gm}$ is the gravitomagnetic flux, $\vec{B}_g$ is the gravitomagnetic field, and $\vec{\Omega}$ is the angular velocity of the reference frame. If the observer measures this additional gravitational field the principle of equivalence holds and he can not distinguish between the elevator and the gravitational field. If he does not observe this effect, the gravitational Larmor theorem is not valid, as a weak field approximation to Einstein's general relativity theory. We will show later that these "induced" non-Newtonian gravitational fields contribute to account for the mechanical energy absorbed (dissipated) by a rotating body during the phase of angular acceleration (deceleration).

Suppose the gravitational Larmor theorem holds, every rotation corresponds to a gravitomagnetic field, which is many orders of magnitude higher than the gravitomagnetic field responsible for the precession forces in the classical Lense-Thirring effect.

*Gravitomagnetic Barnett Effect*

The authors discussed such rotational effect described as the gravitational Barnett effect[6]. In 1915 Barnett[19] observed that a body of any substance set into rotation becomes the seat of a uniform intrinsic magnetic field parallel to the axis of rotation, and proportional to the angular velocity. If the substance is magnetic, magnetization results, otherwise not. This physical phenomenon is referred to as *magnetization by rotation* or as the *Barnett effect*.

If a mechanical momentum with angular velocity $W$ is applied to a substance, it will create a force on the elementary gyrostats (electrons orbiting the nucleus) trying to align them. This is equivalent to the effect of a magnetic field in this substance $B_{equi}$ and we can write:

$$B_{equi} = -\frac{1}{g_l}\frac{2m}{e}\Omega \qquad (17)$$

where $g_l$ is the Landé factor for obtaining the correct gyromagnetic ratio. We can now apply the same argument to the gravitational case and postolate an equivalent gravitomagnetic field $B_{g\ equi}$ which counteracts the mechanical momentum:

$$B_{g\ equi} = -\frac{2}{g}\Omega \qquad (18)$$

For an electron, $g_l=2$ and we see that physical rotation is indeed equivalent to a gravitomagnetic field. From Equation (18) we can compute the gravitomagnetization acquired by the rotating material:

$$\vec{M}_g = \frac{c_g}{m_{0g}}\vec{B}_{g\ equi} = -\frac{c_g}{m_{0g}}\frac{2}{g}\vec{\Omega} \qquad (19)$$

where $\chi_g$ is the garvitomagnetic susceptibility. Taking into account the coupling between gravitation and electromagnetism presented above we can demonstrate the general result:

$$\vec{M}_g = -\frac{\chi}{\mu_0}\frac{2}{g}\left(\frac{m}{e}\right)^2\vec{\Omega} \qquad (20)$$

where χ is the magnetic susceptibility of the material[6]. This indicates that the gravitomagnetic moment associated with the substance will be extremely small. Therefore we can not use this gravitomagnetic moment to induce macroscopic non-Newtonian gravitational fields. However we can show, following our discussion on the equivalence principle, that if the field of rotation in Equation (18) can not be distinguished from gravitomagnetism, it must be a real field which we can use to induce non-Newtonian gravitational fields. The detection of such fields would represent an indirect proof of the existence of the gravitomagnetic Barnett effect.

*Gravitational Poynting Vector and Gravitational Larmor Theorem in Rotating Bodies with Angular Acceleration*

The gravitational Poynting vector, defined as the vectorial product between the gravitational and the gravitomagnetic fields, $\vec{S}_g = \frac{c^2}{4\pi G}\vec{g}\times\vec{B}_g$, provides a mechanism for the transfer of gravitational energy to a system of falling objects (we will consider in the following a cylindrical mass $m$, with radius $a$ and length $\ell$). It has been shown[20] that using the gravitational Poynting vector, the rate at which the kinetic energy of a falling body increases is completely accounted by the influx of gravitational field energy into the body. Applying the gravitational Larmor Theorem (GLT) to a body with angular acceleration. We get that a time varying angular velocity flux will be associated with a non-Newtonian gravitational field proportional to the tangential acceleration. The gravitational electromotive force produced in a gyrogravitomagnetic experiment can be calculated using the gravitational Faraday induction law as given in Equ (16). Together with the GLT expressed through Equation (15) we get

$$\varepsilon_g = \frac{d}{dt}\oiint_\Sigma \vec{\Omega}\cdot d\vec{s} \qquad (21)$$

The induced non-Newtonian gravitational field associated with this gravitational electromotive force is at the surface of the cylinder is:

$$\oint_\Gamma \vec{g}\cdot d\vec{l} = \frac{d}{dt}\oiint_\Sigma \vec{\Omega}\cdot d\vec{s} \qquad (22)$$

$$\vec{g}_q = \frac{1}{2} a\dot{\Omega}\, \hat{e}_q \tag{23}$$

From this non-Newtonian gravitational field and the gravitomagnetic field produced by the rotating mass current, we can compute a gravitational poynting vector

$$\vec{S}_{g\Omega} = \frac{c^2}{4\pi G}\vec{g}_q \times \vec{B}_{g\Omega} = \frac{1}{4\pi}\left(\frac{a}{a+\ell}\right)\Omega\dot{\Omega} m\, \hat{n} \tag{24}$$

which will also provide an energy transfer mechanism to explain how massive bodies acquire rotational kinetic energy when mechanical forces are applied on them[21]. The rate at which the rotational kinetic energy of a body increases (or decreases) due to the application of external mechanical forces on that body, is completely accounted by the influx (out-flux) of gravitational energy into (outward) the body.

$$\left.\frac{dU}{dt}\right|_{\Omega} = S_{g\Omega}(2\pi a^2 + 2\pi a\ell) = \frac{d}{dt}\left(\frac{1}{2} I\Omega^2\right) \tag{25}$$

where $I$, is the moment of inertia of the cylinder. This demonstrates the validity of the gravitational Larmor theorem, and shows how the transfer of mechanical work to a body can be interpreted as a flux of gravitational energy associated with non-Newtonian gravitational fields produced by time varying angular velocities. This is an encouraging result regarding the possible detection of macroscopic non-Newtonian gravitational fields induced through the angular acceleration of the cylinder in the region located outside the rotating cylinder. The non-Newtonian gravitational field outside the cylinder is given by:

$$g = \frac{1}{2}\frac{a^2}{r}\dot{\Omega} \tag{26}$$

where $r > a$ is the distance from the cylinder's longitudinal axis. For $r \leq a$ we have, $g = \frac{1}{2} a\dot{\Omega}$. For the following values of r=1 m, a≈0.1 m, $\dot{\Omega} = 200$ Hz/s, γ will have the value of 1 ms$^{-2}$. We recommend that experiments shall be performed with the aim of evaluating Equation (26).

Is it possible to use fluxes of radiated electromagnetic energy to counteract the effect of absorbed fluxes of gravitational energy? That is a question Saxl, Woodward and Yamashita tried to evaluate empirically. These empirical approaches shall be complemented in the following by a theoretical analysis of the net energy flow associated with the free fall of an electrically charged cylindrical mass.

**Free Fall of a Cylindrical Mass Electrically Charged**

A cylindrical mass electrically charged in free fall must comply with the law of conservation of energy and with the principle of equivalence[22]. During the free fall the cylindrical mass will absorb gravitational energy, which is described by the following gravitational Poynting vector:

$$\vec{S}_g = \frac{c^2}{4\pi G}\vec{g}\times\vec{B}_g = \frac{mv g}{2\pi a\ell}\hat{n}_{in} \qquad (27)$$

where $v$ is the speed of the cylinder while it is falling, $g$ is the Earth gravitational field, $m, a, \ell$ are respectively the mass, length and radius of the cylinder and $\hat{n}_{in}$ is a unit vector orthogonal to the surface of the cylinder and Poynting inwards. The cylinder due to its electric charge will also radiate electromagnetic energy according to the following electromagnetic Poynting vector:

$$\vec{S}_{em} = \frac{1}{\mu_0}\vec{B}\times\vec{E} = \frac{\mu_0}{8\pi^2}\frac{Q^2 v\dot{v}}{a\ell(a+\ell)}\hat{n}_{out} \qquad (28)$$

where $Q$ is the electric charge carried by the cylinder, $\mu_0$ is the magnetic permeability of vacuum and $\hat{n}_{out}$ is a unit vector orthogonal to the surface of the cylinder and Poynting outwards. The principle of equivalence states that if the cylinder is at rest with respect to a reference frame which is uniformly accelerating upwards (with respect to the laboratory) with acceleration $\dot{\vec{v}} = g\hat{e}_z$, the cylinder will radiate (with respect to the laboratory) according to the following Poynting vector:

$$\vec{S}_{em} = \frac{1}{\mu_0}\vec{B}\times\vec{E} = \frac{\mu_0}{8\pi^2}\frac{Q^2 v g}{a\ell(a+\ell)}\hat{n}_{out} \qquad (29)$$

Therefore to comply with the principle of equivalence, we shall take in Equation (28) $\dot{v} = g$ [23].

The sum of both energy fluxes in Equations (27) and (29) must comply with the law of conservation of energy. Therefore the Sum of gravitational incoming flux and the radiated electromagnetic energy flux must be equal to the rate at which the kinetic energy of the body varies in time.

$$S_{em}(2\pi a^2 + 2\pi a\ell) + S_g\, 2\pi a\ell = \frac{d}{dt}\left(\frac{1}{2}mv^2\right) \qquad (30)$$

From Equation (30) we deduce that the acceleration with which the electrically charged cylindrical mass will fall is:

$$\dot{v} = g\left(1 - \frac{m_0}{4\pi}\frac{Q^2}{m\ell}\right) \qquad (31)$$

Equation (31) shows that the free fall of an electrically charged body would violate the law of Galilean free fall, because the acceleration of fall would depend on the electric charge, size and mass of the falling body. The fact that the acceleration of fall depends on the square of the electric charge rules out the possibility to explain with the present analysis, the observations of Saxl and Yamashita, regarding the increase of mass for positively charged bodies and the decrease of mass for negatively charged bodies. Notice that following the rational which leads to equation (31), the phenomenon described by this equation should happen either in a reference frame at rest in an external gravitational field or inside a uniformly accelerated reference frame, therefore we are not able to use this phenomenon to distinguish between both situations. Consequently equation (31) do not violate the principle of equivalence. To test equation (31) we propose to measure the time of fall of charged cylindrical capacitors, and compare it with the time of fall of similar uncharged capacitors. For $m=10$ grams, $l=10$ cm, $Q=100$ C, we will have $\dot{v}=0$. For these values the cylinder would not be able to fall! However to avoid disruption currents for such a high value of electric charge is a technological challenge.

**Conclusion**

In the present work we did an extensive revue of possible "classical ways" to induce non-Newtonian gravitational fields from electromagnetic phenomena or by using the principle of equivalence in the limit of weak gravitational fields. If the experiments performed by Saxl, Yamashita, Woodward and Wallace were reproducible this would represent a breakthrough in the possibility to control gravitational phenomena at the laboratory scale. The understanding of the principle of equivalence for electrically charged bodies and in the limit of weak gravitational fields is crucial to evaluate respectively:

- the possibility of directly convert gravitational energy into electromagnetic energy during the free fall of an electrically charged body.
- the possibility of inducing non-Newtonian gravitational fields through the angular acceleration we might communicate to solid bodies.

The experimental confirmation of such phenomena would be a dramatic step forward in the technological control of free fall. The non detection of the presented phenomena could lead to a better empirical understanding of Einstein's general relativity theory in the limit of weak gravitational fields and when extended to electrically charged bodies, which is a significant scientific result as well. These experiments could also contribute to decide which approach to weak gravity is the correct one, i.e. linearized general relativity or the extension of Newton's laws to time dependent systems.

[23]Soker, N., "Radiation from an Accelerated Charge and the Principle of Equivalence", NASA Breakthrough Propulsion Physics Workshop Proceedings, NASA / CP – 1999-208694, 1999, pp. 427-440